\documentclass[preprint,showpacs,preprintnumbers,amsmath,amssymb]{revtex4}

\usepackage{graphicx}
\usepackage{dcolumn}
\usepackage{bm}

\begin{document}

\title{The Cotton tensor in Riemannian spacetimes}
\author{Alberto A. Garc\'{\i}a}
\email{aagarcia@fis.cinvestav.mx}
\affiliation{Departamento de F\'{\i}sica, CINVESTAV--IPN,\\
Apartado Postal 14--740, C.P. 07000, M\'exico, D.F., M\'exico}
\author{Friedrich W. Hehl}
\email{hehl@thp.uni-koeln.de}
\affiliation{Institute for
Theoretical Physics, University of Cologne,  D--50923 K\"oln,
Germany, and Department of Physics and Astronomy, University of
Missouri-Columbia, Columbia, MO 65211, USA}
\author{Christian Heinicke}\email{
chh@thp.uni-koeln.de}
\affiliation{Institute for Theoretical
Physics, University of Cologne,  D--50923 K\"oln, Germany}
\author{Alfredo Mac\'{\i}as}
\email{amac@xanum.uam.mx}
\affiliation{Departamento de
F\'{\i}sica, Universidad Aut\'onoma Metropolitana--Iztapalapa\\
Apartado Postal 55--534, C.P. 09340, M\'exico, D.F., M\'exico}

\date{20 January 2004}

\begin{abstract}
  Recently, the study of three-dimensional spaces is becoming of great
  interest. In these dimensions the Cotton tensor is prominent as the
  substitute for the Weyl tensor. It is conformally invariant and its
  vanishing is equivalent to conformal flatness. However, the Cotton
  tensor arises in the context of the Bianchi identities and is
  present in {\em any\/} dimension $n$. We present a systematic
  derivation of the Cotton tensor. We perform its irreducible
  decomposition and determine its number of independent components as
  $n(n^2-4)/3$ for the first time.  Subsequently, we exhibit its
  characteristic properties and perform a classification of the Cotton
  tensor in three dimensions. We investigate some solutions of
  Einstein's field equations in three dimensions and of the
  topologically massive gravity model of Deser, Jackiw, and Templeton.
  For each class examples are given. Finally we investigate the
  relation between the Cotton tensor and the energy-momentum in
  Einstein's theory and derive a conformally flat perfect fluid
  solution of Einstein's field equations in three dimensions. {\em
    file cott16.tex, 2004-01-20}
\end{abstract}
\maketitle


\section{Introduction}
The non--linear coupling of gravity to matter in general relativity
presents difficult technical problems in attempts to understand the
gravitational interaction of elementary particles and strings or to
investigate details of the gravitational collapse.  Progress in the
former area has come mainly from treating quantum fields as
propagating on fixed background geometries \cite{pol}, whereas much of
the progress in the latter has come from detailed numerical work
\cite{choptuik,chop,gund}.  Exact solutions of the relevant
matter--gravity equations can play an important role by shedding light
on questions of interest in both general relativity and string theory.
One is often interested in certain classes of solutions with specified
asymptotic properties, the most common of them are the asymptotically
flat spacetimes. Recent work in string theory has, via the AdS/CFT
conjecture, highlighted the importance of the asymptotically anti--de
Sitter spacetimes \cite{malda}. The AdS/CFT correspondence relates a
quantum field theory in $d$ dimensions to a theory in $d+1$ dimensions
that includes gravity \cite{gub,witten}. This is the motivation for
looking at the {\em conformally flat} spaces and at the spaces of {\em
  constant curvature}. For this reason we decided to review the
subject and to collect some old and new results that are nowadays
important in the context of anti--de Sitter spacetimes and to present
them in a modern language. These results seem presently not to be too
well known in the community.

In the theory of conformal spaces the main geometrical objects to be
analyzed are the Weyl \cite{weyl} and the Cotton \cite{cotton}
tensors. It is well known that for conformally flat spaces the Weyl
tensor has to vanish.  Then the Cotton tensor vanishes, too. However,
the Cotton tensor is only conformally invariant in three dimensions.

Recently, the study of three-dimensional spaces is becoming of great
interest; for these spaces the Weyl tensor is always zero and the
vanishing of the Cotton tensor depends on the type of relation between
the Ricci tensor and the energy--momentum tensor of matter.  Any
three-dimensional space is conformally flat if the Cotton tensor
vanishes. If matter is present, the Ricci tensor is related to the
energy--momentum tensor of matter by means of the Einstein equations.
Then the vanishing of the Cotton tensor imposes severe restrictions on
the energy--momentum tensor.  The Cotton tensor also plays a role in
the context of the Hamiltonian formulation of general relativity, see
\cite{adm}.

The outline of the article is as follows. First we derive the Cotton
2-form in the context of the Bianchi identities. Subsequently we
describe its characteristic properties and perform an irreducible
decomposition with respect to the (pseudo-)orthogonal group. This
allows us to determine the number of irreducible components in any
dimension.  Moreover, in four dimensions, we relate the Cotton to the
Bach tensor.  After that we show how to derive the Cotton 2-form in 3
dimensions by means of a variational procedure. We classify the Cotton
2-form in 3 dimensions by means of its eigenvalues and give examples
for all classes.  Finally we derive, in 3 dimensions, the conformally
flat static and spherically symmetric perfect fluid solution of
Einstein's field equation by means of the relation between the Cotton
2-form and the energy--momentum 2-form of matter.

Our notation and conventions are taken from \cite{schouten,PRs}, see
also Appendix \ref{conventions} for an overview.

\section{Bianchi identities and the irreducible decomposition of the
  curvature}

Let $V_n$ be a Riemannian space of $n$ dimensions. We have the {\em
  coframe} $\vartheta^\alpha$ and its dual {\em frame} $e_\alpha$
according to
\begin{equation}
  \vartheta^\alpha = e_i{}^\alpha \, dx^i \,, \quad e_\alpha =
  e^i{}_\alpha \partial_i \,, \quad e_\alpha \rfloor \vartheta^\beta =
  e^i{}_\alpha\,e_i{}^\beta=\delta_\alpha^\beta\,,
\end{equation}
where $x^i$ are local coordinates and the $e^i{}_\alpha$ are the n-bein
coefficients. Greek letters $\alpha,\,\beta,\,\dots=1,\,\dots,n$
denote anholonomic, Latin letters $i,j,\,\dots=1,\,\dots,n$ holonomic
indices. The symbol $\rfloor$ stands for the interior product.

We introduce a metric $g$ according to
\begin{equation}
g = g_{\alpha\beta} \, \vartheta^\alpha \otimes \vartheta^\beta
  = g_{ij} \, dx^i \otimes dx^j \,.
\end{equation}
With the metric at our disposal we also have the Hodge dual operator. Then
we define the $\eta$-basis according to
\begin{equation}
  \eta:={}^\star 1\,,\quad\eta_\alpha:={}^\star\vartheta_\alpha\,,\quad
  \eta_{\alpha\beta}:={}^\star (\vartheta_\alpha\wedge
  \vartheta_\beta)\,, \quad \eta_{\alpha\beta\gamma}:={}^\star
  (\vartheta_\alpha\wedge\vartheta_\beta \wedge
  \vartheta_\gamma)\,,\;\dots
\end{equation}
Furthermore, we equip our manifold with a symmetric connection
($u$ is an arbitrary vector),
\begin{equation}
\nabla_u\,e_\alpha=\Gamma_\alpha{}^\beta(u)\,e_\beta\,,\quad
\Gamma_\alpha{}^\beta(\partial_i)=\Gamma_{i\alpha}{}^\beta\,,
\quad \Gamma_\alpha{}^\beta=\Gamma_{i\alpha}{}^\beta\,dx^i\,,
\end{equation}
with
$d\vartheta^\beta=-\Gamma_\alpha{}^\beta\wedge\vartheta^\alpha$.
In a Riemannian space the connection 1-form can be expressed in terms
of the metric and the object of anholonomity
$\Omega^\alpha:=d\vartheta^\alpha$ and
$\Omega_\alpha:=g_{\alpha\beta}\,\Omega^\beta$,
\begin{equation}\label{connx}
\Gamma_{\alpha\beta}=\Gamma_{i\alpha\beta}\, dx^i
= \frac{1}{2} \, dg_{\alpha\beta} + \left( e_{[\alpha} \rfloor
dg_{\beta]\gamma}\right) \vartheta^\gamma +
e_{[\alpha}\rfloor\Omega_{\beta]}
-\frac{1}{2} \, \left(e_\alpha\rfloor e_\beta \rfloor \Omega_\gamma \right)
\vartheta^\gamma \,.
\end{equation}
The curvature 2-form reads
\begin{equation}\label{curv}
R_\alpha{}^\beta := d\Gamma_\alpha{}^\beta -\Gamma_\alpha{}^\gamma \wedge
\Gamma_\gamma{}^\beta
=\frac{1}{2} \, R_{\mu\nu\alpha}{}^\beta \,\vartheta^\mu\wedge\vartheta^\nu\,.
\end{equation}
Since a metric is given, we can lower the second index. Then the
curvature 2-form is antisymmetric $R_{\alpha\beta}=-R_{\alpha\beta}$.
This can be seen if we choose an orthonormal frame
\begin{equation}
g_{\alpha\beta}={\rm diag}(\underbrace{-1,-1,\dots}_{{\rm ind}}
\underbrace{,1,1,\dots}_{n-{\rm ind}})\,,
\end{equation}
where {\em ind\/}, the number of negative eigenvalues, denotes the index
of the metric.  Then $dg=0$ and, according to (\ref{connx}), the
connection is antisymmetric. In turn, from (\ref{curv}) we can infer
the antisymmetry of $R_{\alpha\beta}$.

The exterior covariant derivative of an $p$-form $X_\alpha{}^\beta$ is
given by
\begin{equation}
D X_\alpha{}^\beta  = d X_\alpha{}^\beta - \Gamma_\alpha{}^\gamma \wedge
X_\gamma{}^\beta + \Gamma_\gamma{}^\beta \wedge X_\alpha{}^\gamma \,.
\end{equation}
The Bianchi identities can be formulated concisely with the help of
this definition.  In a Riemannian space the torsion
$T^\alpha=D\vartheta^\alpha$ vanishes.  Thus, the first Bianchi
identity reads
\begin{equation}\label{firstbia}
0=DT^\alpha=DD\vartheta^\alpha=R_\beta{}^\alpha \wedge \vartheta^\beta\,,
\end{equation}
or, in components,
\begin{equation}
R_{[\alpha\beta\gamma]}{}^\delta = 0\,.
\end{equation}
The first Bianchi identity is a (co-)vector valued 3-form with
\begin{equation}
n \,
\left(\begin{array}{c}n\\3\end{array}\right)=\frac{n^2(n-1)(n-2)}{3!}
\end{equation}
independent components that imposes the same number of
constraint equations on the components of the curvature. Accordingly, in
$n$-dimensions, the curvature 2-form has
\begin{equation}
  \left(\begin{array}{c}n\\2\end{array}\right) \,
  \left(\begin{array}{c}n\\2\end{array}\right) - n \,
  \left(\begin{array}{c}n\\3\end{array}\right) =
  \frac{n^2(n-1)(n+1)}{12}
\end{equation}
independent components. For $n=3$, we have 6 independent components and for
$n=4$ (the case of GR) 20 independent components.

The second Bianchi identity is
\begin{equation}\label{2ndbia}
  DR_{\alpha}{}^\beta = 0 \,,\qquad
  \nabla_{[\lambda}R_{\mu\nu]\alpha}{}^\beta =0\,.
\end{equation}

We now perform the irreducible decomposition of the curvature with respect
to the pseudo-orthogonal group \cite{PRs}:
\begin{equation}
\begin{array}{crcl}
n=1 \qquad\null& R_{\alpha\beta} &=& 0\\
n=2 \qquad\null& R_{\alpha\beta} &=& {\rm Scalar}_{\alpha\beta} \\
n=3 \qquad\null& R_{\alpha\beta} &=& {\rm Scalar}_{\alpha\beta}+
{{\rm Ricci}\!\!\!\!\!\!\!\!\nearrow}\,\,_{\alpha\beta} \\
n\geq 4 \qquad\null& R_{\alpha\beta} &=& {\rm Scalar}_{\alpha\beta}+
{{\rm Ricci}\!\!\!\!\!\!\!\!\nearrow}\,\,_{\alpha\beta}
+{\rm Weyl}_{\alpha\beta}
\end{array}
\end{equation}
\begin{itemize}
\item
The ${\rm Scalar}_{\alpha\beta}$-piece is given by
\begin{equation}
  {\rm Scalar}_{\alpha\beta} := -\frac{1}{n(n-1)} \, R \,
  \vartheta_{\alpha}\wedge\vartheta_{\beta}\,,\quad R:=e_\alpha
  \rfloor {\rm Ric}^{\alpha}\,,\quad {\rm Ric}_\alpha:= e_\beta
  \rfloor R_{\alpha}{}^\beta \,,
\end{equation}
where $R$ is the curvature scalar and ${\rm Ric}_\alpha$ the Ricci
1-form. This piece has 1 independent component and is present in any
dimension $n>1$.  In components we have
\begin{equation}
{\rm Ric}_\alpha={\rm Ric}_{\mu\alpha}\,\vartheta^\mu\,,\quad{\rm
Ric}_{\mu\alpha}=R_{\lambda\mu\alpha}{}^\lambda\,,\quad
R=R_{\lambda\mu}{}^{\mu\lambda}\,,
\end{equation}
and
\begin{equation}
{\rm Scalar}_{\mu\nu\alpha\beta}
=
-\frac{2}{n(n-1)} \, R \, g_{\mu[\alpha}g_{\beta]\nu} \,.
\end{equation}
The {\rm Scalar} piece enjoys the obvious symmetry
\begin{equation}
{\rm Scalar}_{\alpha\beta}\wedge\vartheta^\beta=0\,,\qquad {\rm
Scalar}_{[\mu\nu\alpha]\beta}=0\,.
\end{equation}
\item
From dimension 3 onwards the tracefree Ricci piece comes into play,
\begin{equation}
{{\rm Ricci}\!\!\!\!\!\!\!\!\!\!\nearrow}\,\,\,{}_{\alpha\beta}
:= -\frac{2}{n-2} \, \vartheta_{[\alpha}\wedge{\rm
Ric}\!\!\!\!\!\!\!\nearrow_{\beta]}\,,\quad
{\rm Ric}\!\!\!\!\!\!\!\nearrow_\beta :=
{\rm Ric}_\beta - \frac{1}{n}\, R \,
\vartheta_\beta\,.
\end{equation}
It has $\frac{1}{2}(n+2)(n-1)$ independent components.
In index notation this corresponds to
\begin{equation}
  {{\rm Ricci}\!\!\!\!\!\!\!\!\!\!\nearrow}\,\,\,{}_{\mu\nu\alpha
    \beta} = -\frac{4}{n-2} \, g_{[\mu|\,[\alpha} \, {{\rm
      Ric}\!\!\!\!\!\!\!\!\nearrow}\,{}_{\beta |\,\nu]}\,, \quad {{\rm
      Ric}\!\!\!\!\!\!\!\!\nearrow}\,{}_{\alpha\beta} ={\rm
    Ric}_{\alpha\beta}-\frac{1}{n} \, R \, g_{\alpha\beta} \,.
\end{equation}
If we contract the first Bianchi identity (\ref{firstbia}), we find
\begin{equation}
  0=e_\beta \rfloor (R_\alpha{}^\beta\wedge\vartheta^\alpha) = {\rm
    Ric}_\alpha\wedge\vartheta^\alpha\,,
\end{equation}
since $R_\alpha{}^\alpha=0$ in a Riemannian space. Thus, $ {\rm
  Ric}_{\mu\alpha}\,\vartheta^\mu\wedge\vartheta^\alpha=0$ or
\begin{equation}\label{symsym}
{\rm Ric}_{\alpha\beta}={\rm
    Ric}_{\beta\alpha}\,,
\end{equation}
that is, the Ricci tensor is symmetric. This also implies
\begin{equation}
{{\rm
Ricci}\!\!\!\!\!\!\!\!\!\!\nearrow}\,\,\,{}_{\alpha\beta}
\wedge\vartheta^\alpha=0\,.
\end{equation}
\item Finally, in dimension greater than three, the Weyl 2-form
  emerges according to
\begin{equation}
{\rm Weyl}_{\alpha\beta}:=R_{\alpha\beta}
- {\rm Scalar}_{\alpha\beta}
- {{\rm Ricci}\!\!\!\!\!\!\!\!\nearrow}\,\,_{\alpha\beta}\,.
\end{equation}
From the construction it is clear that the Weyl 2-form is totally
traceless, i.\,e.,
\begin{equation}
e_\alpha \rfloor {\rm Weyl}^{\alpha\beta} = - e_\beta \rfloor {\rm
Weyl}^{\alpha\beta}=0 \,,\quad e_\alpha \rfloor e_\beta \rfloor {\rm
Weyl}^{\alpha\beta}=0\,.
\end{equation}
This property also explains the vanishing of the Weyl 2-form in 3
dimensions.  An arbitrary antisymmetric tensor--valued 2-form
$A_{\alpha\beta} = -A_{\beta\alpha}=A_{\mu\nu\alpha\beta}\,
\vartheta^\mu\wedge \vartheta^\nu/2$ in 3 dimensions has 9 independent
components. The condition $e_\alpha \rfloor A^{\alpha\beta}=0$ results
in 3 one-forms, i.\,e., 9 constraint equations that eventually yield
the vanishing of all components.
\end{itemize}

According to \cite{heinicke}, we can combine ${\rm Scalar}_{\alpha
  \beta}$ and ${{\rm Ricci}\!\!\!\!\!  \!\!\!\nearrow}\,
\,_{\alpha\beta}$,
\begin{equation}\label{ldef1}
  {\rm Scalar}_{\alpha\beta} + {{\rm
      Ricci}\!\!\!\!\!\!\!\!\nearrow}\,\,_{\alpha\beta} =
  -\frac{2}{n-2} \, \vartheta_{[\alpha}\wedge L_{\beta]}\,,
\end{equation}
with
\begin{equation}\label{ldef1*}
  L_\alpha := e_\beta \rfloor R_\alpha{}^\beta - \frac{1}{2(n-1)} \, R
  \, \vartheta_\alpha \,,
\end{equation}
i.e., this sum can be expressed in a coherent way in terms of the
1-form $L_\alpha$.  From ${\rm Scalar}_{\alpha\beta}$ and ${{\rm
    Ricci}\!\!\!\!\!\!\!\!\!\!\nearrow}\,\,\,{}_{\alpha\beta}$ it
inherits the property
\begin{equation}\label{syml}
L_\alpha\wedge\vartheta^\alpha=0\,.
\end{equation}
We may expand $L_\alpha$ in components as
\begin{equation}
L_{\alpha\beta}=L_{\beta\alpha}
= {\rm Ric}_{\alpha\beta} - \frac{1}{2(n-1)} \, R \,
g_{\alpha\beta}\,.
\end{equation}
This tensor is sometimes called {\em Schouten tensor}. Also the names
{\em rho tensor} or $P_{\alpha\beta}$ can be found in the literature.
Then the curvature 2-form can be expressed as
\begin{equation}\label{decomp1}
R_{\alpha\beta}={\rm Weyl}_{\alpha\beta}
-\frac{2}{n-2} \, \vartheta_{[\alpha}\wedge L_{\beta]}
\end{equation}
or, in components,
\begin{equation}\label{decomp2}
{\rm Weyl}_{\alpha\beta\gamma\delta}
=
R_{\alpha\beta\gamma\delta} +
\frac{4}{n-2}\,g_{[\alpha|[\gamma}L_{\delta]|\beta]}\,.
\end{equation}

\section{Cotton 2-form}

By applying the exterior covariant derivative to (\ref{decomp1}), we
obtain the following representation of the second Bianchi identity,
\begin{equation}\label{2ndbia2}
0=DR_{\alpha\beta}=D{\rm Weyl}_{\alpha\beta} + \frac{2}{n-2}
\vartheta_{[\alpha} \wedge C_{\beta]} \,,
\end{equation}
where we encounter the {\em Cotton 2-form}
\begin{equation}\label{cottdef}
  C_{\alpha} := DL_\alpha=\frac
  12\,C_{\mu\nu\alpha}\,\vartheta^\mu\wedge\vartheta^\nu
\end{equation}
or, in components,
\begin{equation}
C_{\alpha\beta\gamma}
=
2\left(\nabla_{[\alpha}{\rm Ric}_{\beta]\gamma} -\frac{1}{2(n-1)} \,
\nabla_{[\alpha}Rg_{\beta]\gamma}\right)\,.
\end{equation}
We perform an irreducible decomposition of the Cotton 2-form with
respect to the Lorentz group. We can use the decomposition for the
torsion, as given in \cite{PRs}, since this is also a vector-valued
2-form. Then we have
\begin{equation}
\begin{array}{rcccccc}
C^\alpha &=& {}^{(1)}C^\alpha &+& {}^{(2)}C^\alpha&+&{}^{(3)}C^\alpha
 \\
&=& {\rm TENCOT}&+&{\rm TRACOT}&+&{\rm AXICOT}\,,\\
\frac{1}{2}\,n^2(n-1) &=& \frac{1}{3}\,n(n^2-4)&+&n&+&\frac{1}{6}\,
n(n-1)(n-2)\,,\label{cotdecomp}
\end{array}
\end{equation}
where
\begin{eqnarray}
{}^{(2)}C^\alpha &:=& \frac{1}{n-1} \, \vartheta^\alpha \wedge
(e_\beta\rfloor C^\beta) \,,\\
{}^{(3)}C^\alpha &:=& \frac{1}{3} \, e^\alpha \rfloor
(C_\beta\wedge\vartheta^\beta)\,,\\
{}^{(1)}C^\alpha &:=& C^\alpha - {}^{(2)}C^\alpha - {}^{(3)}C^\alpha \,,
\end{eqnarray}
or, in components,
\begin{eqnarray}
{}^{(2)}C_{\mu\nu}{}^\alpha&=&-\frac{2}{n-1} \,
\delta^\alpha_{[\mu}\,C_{\nu]\beta}{}^\beta \,,\\
{}^{(3)}C_{\mu\nu}{}^\alpha
&=&\hspace{8pt}\frac{1}{3!} \, C_{[\mu\nu\beta]} \, g^{\alpha\beta} \,,\\
{}^{(1)}C_{\mu\nu}{}^\alpha &=&
C_{\mu\nu}{}^\alpha-{}^{(2)}C_{\mu\nu}{}^\alpha
-{}^{(3)}C_{\mu\nu}{}^\alpha\,.
\end{eqnarray}

TENCOT, TRACOT, and AXICOT are the computer algebra names of the
pieces of the Cotton 2-form, denoting the tensor, the trace, and the
axial pieces, respectively.  The number of independent components of
these pieces is given in the third line of (\ref{cotdecomp}). They
arise as follows: TRACOT corresponds to a scalar-valued 1-form
$C:=e_\alpha \rfloor C^\alpha$ with $n$ independent components. In
general, a (co-)vector-valued 2-form has
\begin{equation}
n \, \left(\begin{array}{c}n\\2\end{array}\right)
=\frac{n^2(n-1)}{2}
\end{equation}
independent components.
AXICOT corresponds to a scalar valued 3-form ($C^\alpha\wedge
\vartheta_\alpha$) and thus has
\begin{equation}
\left(\begin{array}{c}n\\3\end{array}\right)
=\frac{n(n-1)(n-2)}{6}
\end{equation}
independent components.
Thus, TENCOT is left with
\begin{equation}
\frac{n^2(n-1)}{2} - \frac{n(n-1)(n-2)}{6} -n = \frac{n}{3}(n-2)(n+2)
\end{equation}
independent components.

We now show that in a Riemannian space the trace piece (TRACOT)
and the axial piece (AXICOT) vanish.
Hence, only the tensor piece (TENCOT) with its
$n(n^2-4)/3$ independent components survives.
This insight seems to be new. For $n=3$, we have 5
and for $n=4$ (the case of GR) 16 independent components.

In order to see the vanishing of AXICOT, we contract the Cotton 2-form with
the coframe and use (\ref{syml}):
\begin{equation}\label{cotsymm1}
  \vartheta^\alpha \wedge C_\alpha =\vartheta^\alpha \wedge DL_\alpha
  = -D(\vartheta^\alpha \wedge L_\alpha) = 0\,,
\end{equation}
or
\begin{equation}
  C_{[\mu\nu\alpha]}=\frac{2}{3!}\,\nabla_{[\mu}L_{\nu\alpha]} =0\,.
\end{equation}
The second Bianchi identity leads to a vanishing trace of the Cotton 2-form
(TRACOT),
$C= e_\alpha \rfloor C^\alpha=0$.
In order to see this, we contract (\ref{2ndbia2}) twice:
\begin{equation}\label{contrbia}
0=e_\beta \rfloor DR^{\alpha\beta}
= e_\beta \rfloor D{\rm Weyl}^{\alpha\beta} - \frac{n-3}{n-2} \, C^\alpha
-\frac{1}{n-2} \vartheta^\alpha \wedge C \,,
\end{equation}
\begin{equation}
0=e_\alpha \rfloor e_\beta \rfloor DR^{\alpha\beta}
= e_\alpha \rfloor e_\beta \rfloor D{\rm Weyl}^{\alpha\beta}
- 2C = -2C \,,
\end{equation}
or
\begin{equation}
  C_{\alpha\beta}{}^\alpha=\nabla_{\alpha}\left({\rm
      Ric}_\beta{}^\alpha-\frac{1}{2} \, R \,
    \delta_\beta^\alpha\right) =0.
\end{equation}
As we see, the second Bianchi identity relates the derivative of the
Weyl 2-form to the Cotton 2-form,
\begin{equation}\label{bia2}
  e_\beta \rfloor D{\rm Weyl}_{\alpha}{}^{\beta} = \frac{n-3}{n-2} \,
  C_\alpha \,.
\end{equation}
This formula allows us to rewrite the Einstein equation as a Maxwell-like
equation for the Weyl tensor, see \cite{bini}, e.\,g.
The Ricci identity intertwines the derivative of the Cotton 2-form
with the Weyl 2-form,
\begin{eqnarray}
DC_\alpha
&=&
DDL_\alpha = -R_\alpha{}^\beta \wedge L_\beta
=-{\rm Weyl}_\alpha{}^\beta\wedge L_\beta
+ \frac{2}{n-2}\, \vartheta_{[\alpha}\wedge L_{\beta]} \wedge L^\beta
\nonumber \\
&=&
-{\rm Weyl}_\alpha{}^\beta\wedge L_\beta\,.
\end{eqnarray}
Consequently, in 3 dimensions, $C_\alpha$ is a covariantly conserved
$2$-form, with $DC_\alpha=0$. Thus it is a candidate for a conserved
current that can be derived by means of a variational procedure.
The properties of the Cotton tensor are summarized in Table I.

Something similar emerges in 4 dimensions. In Appendix \ref{appendixbach}
it is shown that
\begin{equation}\label{dd*c}
DD{}^\star C_\alpha = -D\left({}^\star{\rm
Weyl}_\alpha{}^\beta\wedge L_\beta\right)\,.
\end{equation}
Thus,
\begin{equation}
B_\alpha := D{}^\star C_\alpha + {}^\star {\rm Weyl}_\alpha{}^\beta \wedge
L_\beta
=: B_\alpha{}^\beta \, \eta_\beta
\end{equation}
or, in components,
\begin{equation}
B_{\alpha\beta}=\nabla^\mu C_{\alpha\mu\beta}+L^{\mu\nu}\,{\rm
Weyl}_{\alpha\mu\beta\nu} \,,
\end{equation}
is a covariantly conserved 3-form:
\begin{equation}\label{DBach}
DB_\alpha=0\qquad (\nabla_\beta B_\alpha{}^\beta=0)\,.
\end{equation}
We recognize the {\em Bach tensor} $B_{\alpha\beta}$
\cite{bach,schouten,penrose,tsantillis}\,.  From the symmetry
properties of $C_\alpha$, $L_\alpha$, and ${\rm Weyl}_{\alpha\beta}$
it follows that
\begin{equation}
B_\alpha\wedge\vartheta^\alpha=0 \quad(B_{\alpha}{}^{\alpha}=0)
\,,\qquad
e_\alpha{}\rfloor B^\alpha=0 \quad(B_{[\alpha\beta]}=0)\,.
\end{equation}
Moreover, it transforms as a conformal density and can be derived from
a variational principle. Since in an conformally flat space the Weyl
and the Cotton tensors vanish, the vanishing of the Bach tensor is
also a necessary (but not sufficient) condition for a four dimensional
space to be conformally flat.

\subsection*{$C_\alpha$ as a variational derivative}

It is well known \cite{deser,mielke} that $C_\alpha$ can be obtained
by means of varying the 3-dimensional Chern-Simons action
\begin{equation}
C_{\rm RR} = -\frac{1}{2} \, \left( \Gamma_\alpha{}^\beta \wedge
d\Gamma_\beta{}^\alpha - \frac{2}{3} \, \Gamma_\alpha{}^\beta \wedge
\Gamma_\beta{}^\gamma \wedge \Gamma_\gamma{}^\alpha \right)
\end{equation}
with respect to the metric keeping the connection fixed.
In order to enforce vanishing torsion
\begin{equation}
T^\alpha := D\vartheta^\alpha = d\vartheta^\alpha - \Gamma_\beta{}^\alpha
\wedge \vartheta^\beta\,
\end{equation}
and vanishing nonmetricity
\begin{equation}
  Q_{\alpha\beta}:=-Dg_{\alpha\beta} = -d g_{\alpha\beta} +
  \Gamma_\alpha{}^\gamma \, g_{\gamma\beta}+\Gamma_\beta{}^\gamma\,
  g_{\alpha\gamma}\,,
\end{equation}
we have to apply Lagrange multipliers.  Then the total Lagrangian
reads
\begin{equation}\label{lag}
  L=C_{\rm RR} + \lambda_\alpha \wedge T^\alpha +
  \lambda^{\alpha\beta} \wedge Q_{\alpha\beta}\,,
\end{equation}
where $\lambda_\alpha$ is a 1-form and
$\lambda^{\alpha\beta}=\lambda^{\beta\alpha}$ a symmetric 2-form.
The corresponding field equations read
(for the explicit calculation see Appendix \ref{vari})
\begin{eqnarray}
  \frac{\delta L}{\delta \lambda_\alpha} &=& T^\alpha =0\,,\\
  \label{varlama} \frac{\delta L}{\delta \lambda^{\alpha\beta}} &=&
  Q_{\alpha\beta}=0\,,\\
\label{vargamm}
\frac{\delta L}{\delta \Gamma_\alpha{}^\beta}
&=&
-R_\beta{}^\alpha - \lambda_\beta \wedge \vartheta^\alpha
+2\lambda_\beta{}^\alpha =0\,,\\
\label{varthet}
\frac{\delta L}{\delta \vartheta^\alpha}
&=&
D\lambda_\alpha =0\,,\\ \label{varg}
\frac{\delta L}{\delta g_{\alpha\beta}}
&=&
D\lambda^{\alpha\beta} =0\,.
\end{eqnarray}
We can solve (\ref{vargamm}) for its symmetric and its antisymmetric parts,
\begin{eqnarray}\label{vargamanti}
  R_{[\alpha\beta]} + \vartheta_{[\alpha}\wedge \lambda_{\beta]} &=&
  0\,,\qquad\\ \label{eq67}
    -R_{(\alpha\beta)}+\vartheta_{(\alpha}\wedge\lambda_{\beta)} + 2
    \lambda_{\alpha\beta} &=&0 \,.
\end{eqnarray}
Because of (\ref{varlama}), the symmetric part
of the curvature vanishes,
\begin{equation}
  0 = DQ_{\alpha\beta} = -DDg_{\alpha\beta} = R_\alpha{}^\gamma \,
  g_{\gamma\beta} + R_\beta{}^\gamma g_{\alpha\gamma} =
  2R_{(\alpha\beta)} \,.
\end{equation}
Thus, by means of (\ref{eq67})
\begin{equation}
\lambda_{\alpha\beta} = -\frac{1}{2} \,
\vartheta_{(\alpha}\wedge\lambda_{\beta)} \,.
\end{equation}
According to (\ref{decomp1}), in 3 dimensions,
$R_{\alpha\beta}=-2\, \vartheta_{[\alpha}\wedge L_{\beta]}$.
We substitute this into (\ref{vargamanti}) and find
\begin{equation}
  \lambda_\beta = 2 \, L_\beta\,,\qquad \lambda_{\alpha\beta} =
  -\vartheta_{(\alpha} \wedge L_{\beta)}\,.
\end{equation}
Eventually,
\begin{equation}
  \frac{1}{2} \, \frac{\delta L}{\delta \vartheta^\alpha} =
  C_{\alpha}\,,\qquad \frac{\delta L}{\delta g_{\alpha\beta}} =
  -\vartheta^{(\alpha}\wedge C^{\beta)} \,,\quad -\frac{2}{n-1} \,
  e_\beta\rfloor \frac{\delta L}{\delta g_{\alpha\beta}} =C^\alpha \,.
\end{equation}
In the presence of matter, the gravitational field equation is given
by $\delta(L+L_{\rm mat})/\delta \vartheta^\alpha=0$.  Hence, the
Cotton 2-form can be coupled to the energy-momentum 2-form of matter
\begin{equation}
\Sigma_\alpha := \frac{\delta L_{\rm mat}}{\delta \vartheta^\alpha} \,.
\end{equation}
This is carried out in the topologically massive gravity model of
Deser, Jackiw and Templeton (DJT) \cite{deser}, where the Lagrangian
(\ref{lag}) is enriched by a Hilbert-Einstein term and a cosmological
term ($\ell$ is the gravitational constant and $\theta$ a
dimensionless coupling constant):
\begin{eqnarray}\label{djtlag}
L_{DJT}&=& \theta \, C_{\rm RR} + V_{\rm HE}+V_\Lambda
+\lambda_\alpha\wedge T^\alpha +
\lambda^{\alpha\beta}\, Q_{\alpha\beta} + L_{\rm mat}
\nonumber\\
&=&\theta \,  C_{\rm RR} -\frac{1}{2\ell}\,R^{\alpha\beta}\wedge
\eta_{\alpha\beta} -\frac{\Lambda}{\ell}\,\eta
+\lambda_\alpha\wedge T^\alpha
+\lambda^{\alpha\beta}\, Q_{\alpha\beta}+L_{\rm mat}\,.
\end{eqnarray}
Then the field equation (\ref{varthet}) reads
\begin{equation}
  G_\alpha + \Lambda\eta_\alpha + \frac{1}{\mu}\, C_\alpha = \ell
  \Sigma_\alpha \,,
\end{equation}
where $G_\alpha=\frac{1}{2} \, \eta_{\alpha\beta\gamma}\wedge
R^{\beta\gamma}$ is the Einstein 2-form and the DJT coupling constant
$1/\mu=-2\theta\ell$.  The model of topologically massive gravity in
Riemann-Cartan space by Mielke and Baekler \cite{mielke} considers
additionally to the Chern-Simons term for the curvature also a
corresponding term for the torsion.  This model includes Einstein,
Einstein-Cartan, and the DJT field equations as limiting cases.

In three dimensions the Cotton tensor arises from the variation of the
topological Chern-Simons term. Recently, a similar procedure was
proposed by Jackiw et al.\ \cite{jackiwpi,guralnik,jackiw} for the
case $n=4$ by starting from the corresponding four-dimensional
topological Lagrangian $(1/2) \, \theta \, R_\alpha{}^\beta\wedge
R_\beta{}^\alpha =\theta \, dC_{RR}$, where $\theta$ is an external,
prescribed field. The variation with respect to the metric yields a
four-dimensional Cotton type tensor that differs from the one in our
definition (\ref{cottdef}) with (\ref{ldef1*}).

The Einstein $(n-1)$-form $G_{\alpha}$ is equivalent to the 1-form
$L_\alpha$ according to
\begin{equation}
G_\alpha = L^\beta\wedge\eta_{\beta\alpha}\,,
\end{equation}
see \cite{heinicke}.  Hence, we may rewrite the DJT-field equation as
an differential equation for $L_\alpha$\,,
\begin{equation}
DL_\alpha + \mu \, L^\beta\wedge \eta_{\beta\alpha}=\ell\mu \,
\Sigma_{\alpha}-\mu\Lambda\eta_\alpha\,.
\end{equation}
The Bianchi identities imply full integrability of this system.

In the case of Einstein gravity, the equivalence of $G_\alpha$ and
$L_\alpha$ implies a relation between the Cotton 2-form and the
energy--momentum $(n-1)$-form in any dimension. We can solve
\begin{equation}\label{eins}
G_\alpha + \Lambda \, \eta_\alpha = \ell \Sigma_\alpha
\end{equation}
for $L_\alpha$ and obtain by covariant exterior differentiation
\begin{equation}\label{cotenergy}
 C_\alpha = (-1)^{n-1+{\rm ind}}\,
\ell \, D^\star \big[\Sigma_\alpha - \frac{1}{n-1} \,
e_{\alpha}\rfloor (\vartheta^\beta\wedge\Sigma_\beta)\big]\,.
\end{equation}
Note that the cosmological constant $\Lambda$, which induces a
constant curvature term, drops out.

We recognize that all vacuum solutions of Einstein's theory have a vanishing
Cotton 2-form. Therefore, via the Bianchi identity, the Weyl 2-form is
divergenceless, see (\ref{bia2}). This property considerably simplifies
the classification of the Weyl tensor.
Petrov type D spacetimes with
vanishing Cotton 2-form have been classified in \cite{ferrando}.

\begin{table}
\caption{Properties of the Cotton 2-form $C_\alpha$ in arbitrary dimensions}
\medskip

\begin{tabular}{|l|l|}\hline
\quad $C_\alpha := DL_\alpha \,,\quad L_\alpha
:= e_\beta \rfloor R_{\alpha}{}^{\beta}
-\frac{1}{2(n-1)} \, R \, \vartheta_\alpha$\hspace{10pt}
&\quad  Cotton 2-form\quad\\
\quad $\vartheta^\alpha \wedge C_\alpha = 0$\qquad (axialfree)
&\quad 1st Bianchi identity\quad \\
\quad $e_\alpha \rfloor C^\alpha =0$\qquad (tracefree)
& \quad contracted 2nd Bianchi identity$\hspace{10pt}$\\
\quad $D{\rm Weyl}_{\alpha\beta}
=-\frac{2}{n-2} \, \vartheta_{[\alpha}\wedge C_{\beta]}$
& \quad 2nd Bianchi identity \quad\\
\quad $DC_\alpha = - {\rm Weyl}_\alpha{}^\beta \wedge L_\beta$
& \quad Ricci identity\quad\\
\quad $\widehat{C}_\alpha = C_\alpha + (n-2) \, \sigma_{,\beta} \, {\rm
Weyl}_\alpha{}^\beta$
&\quad conformal transformation\quad\\
\quad $C_\alpha = \ell \, D^\star [\Sigma_\alpha - \frac{1}{n-1} \,
e_{\alpha}\rfloor (\Sigma_\beta\wedge \vartheta^\beta)] $
&\quad Einstein equation differentiated$\hspace{10pt}$\\
\hline
\end{tabular}
\end{table}

\section{Conformal correspondence}

The conformal correspondence between two $n$--dimensional manifolds
$V_n$ and $\widehat{V}_n$ is achieved by means of a conformal
transformation of the form \cite{Eisen,schouten}
\begin{equation}
  {\hat g}_{\alpha\beta} = \exp(2\sigma) g_{\alpha\beta} \,, \qquad {\hat
    g}^{\alpha\beta} = \exp(-2\sigma) g^{\alpha\beta}\,,
\label{cotr0}\,
\end{equation}
where $\sigma$ is an arbitrary function.  In general, a conformal
transformation (\ref{cotr0}) is {\em not\/} associated with a
transformation of coordinates, i.e., with a diffeomorphism of $V_n$;
both metrics in (\ref{cotr0}) are given in the same coordinate system
and frame.  Since these transformations preserve angles between
corresponding directions, the causal structure of the manifold is
preserved. As a rule, indices of quantities {\em with\/} hat are
raised and lowered by means of $\hat g^{\alpha\beta}$ or $\hat
g_{\alpha\beta}$, respectively, those of untransformed quantities by
$g^{\alpha\beta}$ or $g_{\alpha\beta}$.  The transformed connection
reads
\begin{equation}
  \widehat{\Gamma}_\alpha{}^\beta = {\Gamma}_\alpha{}^\beta + \left(
    \delta_\alpha^\beta \, d\sigma - \vartheta_\alpha \,
    \sigma^{,\beta} + \sigma_{,\alpha} \, \vartheta^\beta \right) =:
  \Gamma_\alpha{}^\beta+S_\alpha{}^\beta\,,
\end{equation}
a comma denotes partial and a semicolon covariant differentiation. If
$\hat D= d + \Gamma_\alpha{}^\beta + S_\alpha{}^\beta$ is the exterior
covariant derivative with respect to
$\widehat{\Gamma}_\alpha{}^\beta$, the transformed curvature is
\begin{equation}\label{curvhat}
\widehat{R}_\alpha{}^\beta
= d\widehat{\Gamma}_\alpha{}^\beta
-\widehat \Gamma_\alpha{}^\gamma\wedge\widehat\Gamma_\gamma{}^\beta
= R_\alpha{}^\beta + 2 \, \vartheta_{[\alpha}\wedge S_{\gamma]} \,
g^{\gamma\beta}\,,\quad
\end{equation}
with
\begin{equation}\label{strans}
S_{\gamma} := D\sigma_{,\gamma} - \sigma_{,\gamma} \, d\sigma + \frac{1}{2} \,
\sigma^{,\alpha}\sigma_{,\alpha} \, \vartheta_\gamma \,.
\end{equation}
By contracting (\ref{curvhat}) with the frame $e_\beta\rfloor$, we
infer
\begin{eqnarray}\label{ltrans}
\widehat{L}_\alpha &=& L_\alpha - (n-2) \, S_\alpha \,,\\
\widehat{\rm Weyl}_\alpha{}^\beta &=& {\rm Weyl}_\alpha{}^\beta\,,\\
\widehat{R}&=&\exp(-2\sigma) \left[R-2(n-1) \, \sigma^{,\alpha}{}_{;\alpha}
- (n-1)(n-2) \sigma_{,\alpha}\sigma^{,\alpha}\right]\,.
\end{eqnarray}
The Weyl 2-form is conformally invariant since a conformal transformation
does not act on the trace-free part of the curvature. Application of
$\widehat{D}$ onto (\ref{ltrans}) yields the transformation behavior
of the Cotton 2-form,
\begin{equation}
\widehat{C}_\alpha = C_\alpha + (n-2) \, \sigma_{,\beta} \, {\rm
Weyl}_\alpha{}^\beta \,.
\end{equation}
Thus, in $n=3$, where the Weyl 2-form vanishes, the Cotton 2-form becomes
conformally invariant.

\section{Criteria for conformal flatness}

In the following paragraphs we investigate the criteria for conformal
flatness, i.\,e., the possibilities to transform the curvature to zero
by means of a conformal transformation. We basically follow
\cite{schouten}.  Since we have seen that the curvature 2-form in 2,
3, and more than 3 dimensions is built up rather differently, we have
to investigate these cases separately.

{$\bf n=2$}

In $n=2$ the only non-vanishing curvature piece is the curvature scalar $R$.
Its behavior under conformal transformation is given by
\begin{equation}
\widehat{R} = \exp(-2\sigma) \, \left( R- 2 \, \sigma^{,i}{}_{;i}\right) = 0
\,.
\end{equation}
Thus,
\begin{equation}
  \widehat{R} =0 \quad \Longleftrightarrow \quad \sigma^{,i}{}_{;i} =
  \frac{R}{2} \,.
\end{equation}
This is a scalar wave equation for the conformal factor $\sigma$
with $R$ as source. Since the wave equation always has a solution, we
conclude that all 2-dimensional spaces are conformally flat.

{$\bf n\geq 3$}

For more than 2 dimensions we start from (\ref{decomp1}), namely
\begin{equation}
R_{\alpha\beta} ={\rm Weyl}_{\alpha\beta}
-\frac{2}{n-2} \, \vartheta_{[\alpha}\wedge L_{\beta]} \,.
\end{equation}
Since the Weyl 2-form is conformally invariant it cannot be
transformed to zero by means of a
conformal transformation. Consequently, the vanishing of the Weyl 2-form is
a necessary condition for conformal flatness.

The $L_\alpha$ 1-form transforms according to
\begin{equation}
\widehat{L}_\alpha = L_\alpha - (n-2) \, S_\alpha \,.
\end{equation}
We can transform $L_\alpha$ to zero if there is a function $\sigma$ such
that
\begin{equation}
L_\alpha = (n-2) \, S_\alpha \,.
\end{equation}
This will impose a differential restriction on  $L_{ij}$. By means of
(\ref{strans}), we rewrite the
latter equation as a differential equation for
$\sigma_{,i}$,
\begin{equation}\label{dsigma}
D\sigma_{,i}=\sigma_{,i} \, \sigma_{,j} \, \vartheta^j - \frac{1}{2} \,
\sigma^{,j}\,\sigma_{,j} \, \vartheta_i + \frac{1}{n-2} \, L_i\,.
\end{equation}
If we apply the covariant derivative to both sides of (\ref{dsigma}),
we obtain a necessary condition for the integrability,
\begin{equation}\label{intcond}
-R_i{}^j \, \sigma_{,j} = DD\sigma_{,i}
= \sigma_{,j} \, D\sigma_{,i} \wedge \vartheta^j
- \sigma^{,j} \, D\sigma_{,j} \wedge \vartheta_i + \frac{1}{n-2} \, C_i \,.
\end{equation}
This becomes a necessary and sufficient condition of integrability if
the dependence on $\sigma_{,i}$ can be eliminated, see
\cite{schouten,schouten58}.  Thus we substitute $D\sigma_{,i}$ from
(\ref{dsigma}) into (\ref{intcond}):
\begin{equation}
-R_i{}^j \, \sigma_{,j} = -\frac{2}{n-2} \, L_{[i}\wedge\vartheta_{j]} \,
\sigma^{,j} + \frac{1}{n-2} \, C_i\,.
\end{equation}
Using the decomposition (\ref{decomp1}) of the curvature,
we finally arrive at
\begin{equation}
-(n-2) \, {\rm Weyl}_i{}^j \, \sigma_{,j} = C_i\,.
\end{equation}

For $n=3$, the Weyl 2-form is zero and $C_\alpha = 0$ is the
integrability condition for the conformal factor. Thus, if the Cotton
2-form is zero, the space is conformally flat. Conversely, if the
space is conformally flat, there is a conformal transformation such
that $\widehat R_{\alpha}{}^\beta=0 \Leftrightarrow
\widehat{L}_\alpha=0 \Rightarrow \widehat C_\alpha =0$. Since the
Cotton 2-form is conformally invariant in 3 dimensions, we find
$C_\alpha=0$. Hence, the vanishing of the Cotton 2-form is the
necessary and sufficient condition for a $V_3$ to be conformally flat.

In more than 3 dimensions the vanishing of the Weyl 2-form is a
necessary condition for conformal flatness. Thus, also in dimensions
greater than 3, $C_\alpha = 0$ is the integrability condition for the
conformal factor.  However, for $n>3$, the contracted second Bianchi
identity (\ref{contrbia}) implies the vanishing of the Cotton 2-form
when the Weyl 2-form is zero. Hence, the vanishing of the Weyl 2-form
is also the sufficient condition for conformal flatness.

\section{Classification of the Cotton 2-form in 3D}
\label{class}

A vector-valued 2-form in 3 dimensions has 9 independent components, the
same as the number of components of
a $3\times 3$ matrix. A mapping between these two can be
achieved by means of the Hodge dual. The Hodge dual of a vector-valued
2-form in 3 dimensions is a vector-valued 1-form with the same number of
independent components. Its components form a 2nd rank
tensor (``matrix''),
\begin{equation}\label{cottonyork}
C_{\alpha\beta} := e_\alpha \rfloor {}^\star C_\beta = {}^\star
(C_\beta\wedge \vartheta_\alpha)
\end{equation}
or, in components,
\begin{equation}\label{cottonyork2}
C_\alpha{}^\beta=\nabla_\mu\left({\rm
Ric}_{\nu\alpha}-\frac{1}{4}Rg_{\nu\alpha}\right) \, \eta^{\mu\nu\beta}\,.
\end{equation}
This alternative representation of the Cotton 2-form, often called
{\em Cotton-York tensor} \cite{york} (even though it was already
discussed explicitly by ADM \cite{adm}), can only be defined in three
dimensions.  Sometimes it appears under the name Bach tensor in the
literature, see \cite{christo}, e.g. This seems to be a misnomer.

The Cotton tensor is tracefree
\begin{equation}
C_\alpha{}^\alpha = e_\alpha \rfloor {}^\star C^\alpha =
{}^\star(C^\alpha\wedge \vartheta_\alpha) = 0 \,.
\end{equation}
In the 3 dimensions, the 2nd Bianchi identity (\ref{2ndbia2}) amounts to
$\vartheta_{[\alpha} \wedge C_{\beta]} = 0$. In view of the definition
(\ref{cottonyork}), we infer that the Cotton tensor is symmetric
$C_{\alpha\beta}=C_{\beta\alpha}$.
Introducing this symmetry explicitly into (\ref{cottonyork2}), we obtain
the alternative representation
\begin{equation}
C^{\alpha\beta}=C^{\beta\alpha}= \eta^{\mu\nu(\alpha}\nabla_\mu {\rm
Ric}_{\nu}{}^{\beta)}\,.
\end{equation}

We now perform a classification of the Cotton tensor with respect to
its eigenvalues.
 The corresponding generalized eigenvalue problem reads:
\begin{equation} \label{eigen2}
\left(C^{\alpha\beta}-\lambda\,g^{\alpha\beta}\right)
V_\beta=0\,,\quad C^{[\alpha\beta]}=0\,,\quad
 C^{\alpha\beta}\,g_{\alpha\beta}=0\,.
\end{equation}
By lowering one index, we can reformulate this as ordinary eigenvalue
problem for the matrix $C_\alpha{}^\beta$. However, in that case,
the symmetry $C^{\alpha\beta}=C^{\beta\alpha}$ is no longer manifest:
\begin{equation} \label{eigen1}
\left(C_\alpha{}^\beta-\lambda\,\delta_\alpha^\beta\right)
V_\beta=0\,,\quad C_\alpha{}^\alpha=0\,.
\end{equation}

\subsection{Euclidean signature}

The case of Euclidean signature is simple: the generalized eigenvalue
problem reduces to an ordinary one. As a real symmetric matrix
$C^{\alpha\beta}$ possesses 3 real eigenvalues and the eigenvectors
form a basis. With respect to this basis, $C^{\alpha\beta}$ takes a
diagonal form. Since $C^{\alpha\beta}$ is tracefree, the sum of the
eigenvalues is zero. Consequently, we can distinguish 3 classes:
\begin{itemize}
\item Class A\\
Three distinct eigenvalues: $\lambda_1 \neq \lambda_2$ and
$\lambda_3=-(\lambda_1+\lambda_2)$.
\item Class B\\
Two distinct eigenvalues: $\lambda_1=\lambda_2\neq 0$,
$\lambda_3=-2\lambda_1$.
\item Class C\\ One distinct eigenvalue:
  $\lambda_1=\lambda_2=\lambda_3=0$.\\ In the present context of
  Euclidean signature, this implies $C_{\alpha\beta}=0$.

\end{itemize}

\subsection{Lorentzian signature}
In the case of an indefinite metric, the roots of the characteristic
polynomial
\begin{equation}
\det \left(C^{\alpha\beta}-\lambda \, g^{\alpha\beta}\right)=0
\end{equation}
may be complex. Accordingly, the matrix $C_\alpha{}^\beta$ is no
longer symmetric and in the equivalent ordinary eigenvalue problem
\begin{equation}
\det\left(C_\alpha{}^\beta -\lambda \, \delta_\alpha^\beta\right) = 0
\end{equation}
complex eigenvalues occur, too.  This point seems to have been
overlooked by the authors of \cite{barrow}.  Consequently, the
classification will not be as simple as it was the case for the
Euclidean metric.

In the following, we will present a classification of
$C_\alpha{}^\beta$. The tracefree condition (\ref{eigen1})$_2$, in
orthonormal coordinates, reads explicitly
\begin{equation}\label{}
C_1{}^1+C_2{}^2+C_3{}^3=0\,.
\end{equation} Accordingly, we can eliminate $C_3{}^3$,
e.g., from (\ref{eigen1})$_1$. Then the secular determinant reads
\begin{equation}\label{}
  {\rm det}\left|\begin{array}{ccc}C_1{}^1-\lambda & C_1{}^2 &
      C_1{}^3\\ -C_1{}^2&C_2{}^2-\lambda& C_2{}^3\\-C_1{}^3&
      C_2{}^3&-C_1{}^2-C_2{}^2-\lambda
\end{array}\right| =0\,,
\end{equation}
with the 5 matrix elements $C_1{}^2,C_1{}^2,C_1{}^3,C_2{}^2,C_2{}^3$.
We compute the determinant and order according to powers of $\lambda$,
\begin{equation}\label{charpoly}
\lambda^3+b\,\lambda+c=0\,,
\end{equation}
where
\begin{eqnarray}
  b&:=&-(C_1{}^1)^2-C_1{}^1C_2{}^2
  -(C_2{}^2)^2+(C_1{}^2)^2+(C_1{}^3)^2-(C_2{}^3)^2 \,, \\
  c&:=&\left[(C_1{}^1)^2C_2{}^2+C_1{}^1(C_2{}^2)^2+C_1{}^1
    (C_1{}^2)^2+C_1{}^1(C_2{}^3)^2 \right.\\ \nonumber & &\qquad
  \left.+(C_1{}^2)^2C_2{}^2+2\,C_1{}^2C_1{}^3C_2{}^3-(C_1{}^3)^2C_2{}^2
  \right]\,.
\end{eqnarray}
The roots of (\ref{charpoly}) are given by
\begin{equation}
\lambda_1 = A\,,\quad \lambda_2 = -\frac{A}{2} \, + i\frac{\sqrt{3}}{2} \, B
\,,\quad \lambda_3=-\frac{A}{2} - i\frac{\sqrt{3}}{2} \, B \,,
\end{equation}
with
\begin{equation}
A:=\frac{D^2-12b}{6D}\,,\quad B:=\frac{D^2+12b}{6D} \,,\quad D:=
\left(-108c+12\sqrt{12b^3+81c^2}\right)^{1/3}\,.
\end{equation}
A cubic polynomial with real coefficients has at least one real root
and the complex roots have to be complex conjugates. The Jordan normal
forms of the Cotton tensor read:
\bigskip

\footnotesize\hspace{-1cm}
\begin{tabular}{cccc}
\mbox{\normalsize ``Petrov''-type}&\qquad
\mbox{\normalsize Jordan form} & \qquad
\mbox{\normalsize Segre notation}&
\mbox{\normalsize eigenvalues}
\nonumber \vspace{10pt}\\
I&$\left(\begin{array}{ccc}
\hspace{8pt}\lambda_1&0&0\\0&\hspace{8pt}
  \lambda_2&0\\0&0&-\lambda_1-\lambda_2\\
\end{array}\right)$ & [111] &
$\lambda_1\neq\lambda_2,\,\lambda_3=-\lambda_1-\lambda_2$\vspace{10pt}\\
D&$\left(\begin{array}{ccc}
\hspace{8pt}\lambda_1&0&0\\0&\hspace{8pt}\lambda_1&0\\0&0&-2\lambda_1\\
\end{array}\right) $& [(11)1] &$\lambda_1=\lambda_2\neq 0,
\lambda_3=-2\lambda_1$\vspace{10pt}\\
II&$\left(\begin{array}{ccc}
\hspace{8pt}\lambda_1&1&0\\0&\hspace{8pt}\lambda_1&0\\0&0&-2\lambda_1\\
\end{array}\right)$ & [21] &$\lambda_1=\lambda_2\neq 0, \lambda_3=
-2\lambda_1$\vspace{10pt}\\
N&$\left(\begin{array}{ccc}
\hspace{8pt}0&\hspace{8pt}1&\hspace{8pt}0$\hspace{8pt}$\\
\hspace{8pt}0&\hspace{8pt}0&
\hspace{8pt}0$\hspace{8pt}$\\
\hspace{8pt}0&\hspace{8pt}0&\hspace{8pt}0$\hspace{8pt}$ \\
\end{array}\right)$ & [(21)] &$\lambda_1=\lambda_2=\lambda_3=0$\vspace{10pt}\\
III&$\left(\begin{array}{ccc}
\hspace{8pt}0&\hspace{8pt}1&\hspace{8pt}0$\hspace{8pt}$\\
\hspace{8pt}0&\hspace{8pt}0&\hspace{8pt}1$\hspace{8pt}$
\\ \hspace{8pt}0&\hspace{8pt}0&\hspace{8pt}0$\hspace{8pt}$\\
\end{array}\right)$ & [3] &$\lambda_1=\lambda_2=\lambda_3=0$\vspace{10pt}\\
O&$\left(\begin{array}{ccc}
\hspace{8pt}0&\hspace{8pt}0&\hspace{8pt}0$\hspace{8pt}$\\
\hspace{8pt}0&\hspace{8pt}0&\hspace{8pt}0$\hspace{8pt}$\\
\hspace{8pt}0&\hspace{8pt}0&\hspace{8pt}0$\hspace{8pt}$\\
\end{array}\right) $& \\
\end{tabular}
\normalsize
\bigskip

This parallels exactly the Petrov classification of the Weyl
tensor in 4 dimensions \cite{kramer}. This comes about since the
Weyl tensor in 4D is equivalent to a (complex) $3\times 3$
tracefree matrix, as $C_\alpha{}^\beta$ in 3D; for a similar
classification of $C_{\alpha\beta}$, see \cite{hall}.

Since one eigenvalue is real, types D and II with only one independent
eigenvalue $\lambda_1=\lambda_2=-2\lambda_3$ are always real. For
class I, besides the real eigenvalue, two complex eigenvalues may
occur. In that case, they are complex conjugated.
Therefore, class I can be subdivided into class I with 3 real eigenvalues,
$[111]$, and class I' with one real and two complex conjugated eigenvalues,
$[1z\bar z]$. By performing a kind of null rotation, we can also give a real
form for class I':\bigskip

\begin{tabular}{cccc}
  I'\qquad\qquad&$\left(\begin{array}{ccc} {\rm Re}\, z&{\rm Im}\,
      z&0\\-{\rm Im}\, z&{\rm Re}\, z&0\\0&0&-2\, {\rm Re}\, z\\
\end{array}\right)$ \qquad \qquad& $[1z\bar z]$\qquad\qquad
&$\lambda_1 =-2\, {\rm Re}\, z,\, \lambda_2=z,\,\lambda_3=\bar z$ \,.
\end{tabular}
\bigskip

We can now specify simple criteria for deciding to which of these
classes the Cotton tensor $C_\alpha{}^\beta$ belongs. First determine
the eigenvalues.
\begin{enumerate}
\item Three different eigenvalues (2 independent)
\begin{enumerate}
\item all real $\Rightarrow$ Class I
\item one real, two complex $\Rightarrow$ Class I'
\end{enumerate}
\item Two different eigenvalues (1 independent
  $\lambda_1=\lambda_2=-2\lambda_3$)
\begin{enumerate}
\item $(C_\alpha{}^\beta - \lambda_1 \, \delta_\alpha^\beta)
(C_\beta{}^\gamma + \frac{1}{2}\, \lambda_1 \, \delta_\beta^\gamma)=0$
$\Rightarrow$ Class D
\item else $\Rightarrow$ Class II
\end{enumerate}
\item All eigenvalues zero
\begin{enumerate}
\item $C_\alpha{}^\beta=0$ $\Rightarrow$ 0
\item $C_\alpha{}^\beta \, C_\beta{}^\gamma =0$ $\Rightarrow$ Class N
\item else $\Rightarrow$ Class III
\end{enumerate}
\end{enumerate}

\section{Examples}
We now give examples in order to demonstrate explicitly that all classes
presented are non-empty indeed. All results have been checked by means of
computer algebra, see Appendix \ref{computeralgebra} for an explicit sample
program.

\begin{itemize}
\item Class I'\\
The generic example is the $(1+2)$D static and spherically symmetric
spacetime, given in an orthonormal coframe with signature $(+--)$ by
\begin{equation}
\vartheta^{\hat 0} = \sqrt{\psi} \, dt,\quad \vartheta^{\hat
1}=\frac{dr}{\sqrt{\psi}}, \quad
\vartheta^{\hat 2} = r \, d\varphi \,, \quad \psi=\psi(r) \,.
\end{equation}
The Cotton tensor and its eigenvalues read, here $()'=d/dr$:
\begin{equation}
C_\alpha{}^\beta = \frac{\sqrt{\psi}\,\psi'''}{4} \, \left(\begin{array}{ccc}
0&0&-1\\0&0&0\\1&0&0\end{array}\right)\,,\quad
\lambda_1=0\,,\lambda_2=-\lambda_3=i\frac{\sqrt{\psi} \, \psi'''}{4} \,.
\end{equation}
A well-known example is the 3D analog to the Reissner-Nordstr\"om solution,
a solution of the 3D Einstein-Maxwell equation \cite{btz}:
\begin{equation}
\psi=\Lambda \, r^2 -q^2\, \ln r -M\,.
\end{equation}

\item Class I\\
In \cite{nutkubaekler}, {eq.(4.1)},
 the following solution for the vacuum DJT
field equation is given:
The orthonormal coframe with signature $(-++)$ reads
\begin{eqnarray}
\vartheta^{\hat 0} &=& a_0 \,(d \psi+\sinh \theta \, d\phi) \,, \\
\vartheta^{\hat 1} &=& a_1 \, (-\sin\psi \, d \theta+\cos\psi \,
\cosh\theta \, d\phi)\,,\\
\vartheta^{\hat 2} &=& a_2 \ (\cos\psi \, d\theta + \sin\psi \cosh\theta \,
d\phi) \,,
\end{eqnarray}
where the DJT field equations are fulfilled provided
\begin{equation}
a_0+a_1+a_2=0\,,\qquad \mu= -\frac{a_0^2+a_1^1+a_2^2}{a_0 a_1 a_2}\,.
\end{equation}
Then the Cotton tensor reads
\begin{equation}
C_\alpha{}^\beta = -4\frac{a_1^2+a_1a_2+a_2^2}{(a_1+a_2)a_1^2a_2^2}
\, \left(\begin{array}{ccc}
1&0&0\\0&\frac{-a_1}{a_1+a_2}&0\\0&0&\frac{-a_2}{a_1+a_2}
\end{array}\right).
\end{equation}
The eigenvalues can be read off from the diagonal. For $a_1=a_2$, the
Cotton tensor degenerates to class D. The solution eq.(4.6) in
\cite{nutkubaekler} is analogous to the present case.

\item Class D\\
An example is the 3D G\"odel solution (signature $(+--)$), see
\cite{vurio} {eq.(4.1)}:
\begin{eqnarray}
\vartheta^{\hat 0} &=& \left(\frac{3}{\mu}\right) \, \left[(dt - 2(\sqrt{r^2
+1} - 1) \, d\phi\right]\,,\\
\vartheta^{\hat 1} &=&\left(\frac{3}{\mu}\right) \,\frac{dr}{\sqrt{r^2+1}}\,, \\
\vartheta^{\hat 2} &=&\left(\frac{3}{\mu}\right) \,r\, d\phi \,,
\end{eqnarray}
with
\begin{equation}
C_\alpha{}^\beta = \left(\frac{\mu}{3}\right)^3 \, \left(\begin{array}{ccc}
-2&0&0\\0&1&0\\0&0&1\end{array}\right)\,,\quad
\lambda_1=\lambda_2=-\frac{1}{2}\lambda_3=\left(\frac{\mu}{3}\right)^3 \,.
\end{equation}
This is a vacuum solution of the DJT model as well as a solution of the 3D
Einstein equation with matter.

Many of the other solutions known for the DJT field equation are also
of Class~D:
\begin{itemize}

\item The squashed 3-sphere solutions by Nutku and Baekler
\cite{nutkubaekler}, eq.(4.10) and eq.(4.1), eq.(4.6) for a special choice
of parameters (see above).

\item The topologically massive planar universe with constant twist of
Percacci et al.\ \cite{vurio}, eq.(3.20).

\item The perfect fluid solution of G\"urses \cite{gurses}, eq.(6).

\item The DJT-black hole solution of Nutku \cite{nutku}, eq.(24).

\item The recent black hole solution by Moussa et al.\
\cite{clement}, {eq.(4)}.

\end{itemize}

\item Class N\\
Inspired by the corresponding $(1+3)$D metrics, we start with the ansatz
\begin{equation}
\vartheta^{\hat 0}=dt+dx\,,\quad
\vartheta^{\hat 1}=dt- dx\,,\quad
\vartheta^{\hat 2}=dy\,,
\end{equation}
with the {\em non-orthonormal} metric
\begin{equation}
g=\vartheta^{\hat 0}\otimes\vartheta^{\hat 1}
+ \psi\,\vartheta^{\hat 1}\otimes\vartheta^{\hat 1}
- \vartheta^{\hat 2} \otimes \vartheta^{\hat 2}\,,\quad \psi=\psi(y)\,.
\end{equation}
The Cotton tensor, in this frame, reads (\,$()'=d/dy$):
\begin{equation}
C_\alpha{}^\beta = \psi''' \, \left(\begin{array}{ccc}
0&0&0\\-1&0&0\\0&0&0\end{array}\right)\,,\quad
\lambda_1=\lambda_2=\lambda_3=0 \,.
\end{equation}
The vacuum DJT field equation reduces to
\begin{equation}
\frac{1}{\mu}\, \psi'''-\psi''=0\,,
\end{equation}
with the general solution
\begin{equation}
\psi=Ay+B\, e^{\mu y} + C\,.
\end{equation}
In an orthonormal coframe with signature $(-++)$ and $A=C=0$ and $B=1$,
coframe and Cotton tensor can be brought into the more familiar form
\begin{eqnarray}
\vartheta^{\hat 0}& = &e^{\mu y/2} \, \left[ (1+\frac{1}{2} \, e^{-\mu y}) \,
                       dt
                   + (1-\frac{1}{2} \, e^{-\mu y}\, dx)\right] \,,\\
\vartheta^{\hat 1} &=& \frac{1}{2} \, e^{-\mu y/2} \, (dt - dx) \,,\\
\vartheta^{\hat 2} &=& dy \,.
\end{eqnarray}
The Cotton tensor, with all eigenvalues being zero, reads
\begin{equation}
C_\alpha{}^\beta = \frac{\mu^3}{2} \, \left(\begin{array}{ccc}
-1&-1&0\\1&1&0\\0&0&0\end{array}\right)\,,\quad
\lambda_1=\lambda_2=\lambda_3=0  \,.
\end{equation}

Another class N solution is given in \cite{vurio}, eq.(4.9).
\end{itemize}

We have found no (sensible) solutions to the Einstein or DJT field
equations which are of Class II or III. However, it is easy to find
metrics for which the Cotton tensor is in general of class I but may
degenerate to classes II or III. Just in order to show that these
classes are nonempty, we will sketch corresponding examples:
\begin{itemize}
\item Class II\\
The following coframe (signature $(-++)$),
\begin{equation}
\vartheta^{\hat 0}=e^{-2y} \, dt + dx\,,\quad
\vartheta^{\hat 1}=e^{y} \, dx\,,\quad
\vartheta^{\hat 2}=dy\,,
\end{equation}
yields a Cotton tensor which, in general, is of class I. However, on
the surface $y=0$, it degenerates to class II.

\item Class III\\
The Cotton tensor for the following coframe (signature $(-++)$)
is also of class I in general:
\begin{equation}
\vartheta^{\hat 0} = (x-t) \, dt\,,\quad
\vartheta^{\hat 1} = (x+t) \, dx \,,\quad
\vartheta^{\hat 2} = dy \,.
\end{equation}
On the hypersurface given by $x=t(\sqrt{13}+3)/2$, the Cotton
tensor degenerates to class III.

\item Class 0\\
All conformally flat solutions.
\end{itemize}

\subsection*{Conformally flat perfect fluid solution}
As an application of the relation between energy-momentum 2-form and
Cotton 2-form and as an example for a class 0 solution, we will
derive the spherically symmetric, conformally flat, perfect fluid
solution to Einstein's field equation.  We use the ansatz
\begin{equation}\label{ansatz}
\vartheta^{\hat 0} = N(r) \, dt\,,\quad
\vartheta^{\hat 1} = dr/F(r)\,,\quad
\vartheta^{\hat 2} = r \, d\phi\,,
\end{equation}
with signature $(-++)$. The energy--momentum of the perfect fluid is
given by
\begin{equation}
\Sigma_\alpha = [\rho(r)+p(r)] \, u_\alpha \, u^\beta\eta_\beta +
p\,\eta_\alpha \,,
\end{equation}
where $u^\alpha$ is the 4-velocity of the fluid elements
which, in an orthonormal coframe, is given by $u^\alpha=(1,0,0)$.
By using (\ref{cotenergy}), we find
\begin{eqnarray}
  C_{\hat 0} \label{cotfluid0} &=& -\left\{\frac{F}{2N} \, \left(2\,
      \partial_r \,[N \, (p+\rho)] - N \, \partial_r
      \rho\right)\right\} \, \vartheta^{\hat 0} \wedge \vartheta^{\hat
    1} \,,\\ C_{\hat 1} &=& 0\,,\\ C_{\hat 2} &=& -\frac{F}{2} \,
  \partial_r \rho \, \vartheta^{\hat 1}\wedge\vartheta^{\hat 2}\,.
\end{eqnarray}
Consequently, we have to demand constant energy density $\rho=const$ for a
conformally flat solution with $C_\alpha=0$. By using $\rho=const$, we infer
from (\ref{cotfluid0})
\begin{equation}\label{nn}
N(r) = \frac{c_1}{\rho+p(r)} \,,
\end{equation}
where $c_1$ is an integration constant.
The $0$-component of the Einstein field equation (\ref{eins})
yields
\begin{equation}\label{ff}
F^2(r)=c_2 - (\ell\rho+\Lambda)\,r^2\,.
\end{equation}
The remaining components of the field equation are fulfilled provided
\begin{equation}
\frac{dp}{dr}= \frac{(\ell\rho-\Lambda)\,(p+\rho)\,r}{F^2}\,.
\end{equation}
This ordinary differential equation can be integrated yielding ($c_3$ is
another integration constant)
\begin{equation}\label{pp}
p=\frac{c_3 \, F \, (\ell\rho+\Lambda)+ (c_3)^2 \ell \Lambda + \rho
F^2}{(c_3)^2
\, \ell^2 - F^2}\,.
\end{equation}
Finally, the solution is given by (\ref{ansatz},\ref{nn},\ref{ff},\ref{pp}),
compare the solutions in \cite{camp}.

\begin{acknowledgments}
Discussion with Yuri Obukhov (Cologne) are gratefully
acknowledged. This research was supported by CONACyT Grants:
42191--F, 38495--E,  and by the joint German--Mexican project
CONACYT --- DFG: E130--655 --- 444 MEX 100 and by DFG project HE
528/20-1.
\end{acknowledgments}

\section{Appendix}

\subsection{Conventions}\label{conventions}

Our index notation is based on the conventions of Schouten
\cite{schouten} and for exterior calculus we refer to \cite{PRs}.  For
quick and easy reference, we display our conventions for index
positions and signs of the Christoffel symbol, the Riemann tensor, and
the Ricci tensor (holonomic indices $i,j,\dots =0,1,2,3$).  The sign
of the Ricci tensor is the same as those of the $L_{ij}$ tensor and
the Cotton tensor. In particular, the ${\rm Ric}_{ij}$ sign introduces
a relative sign between the $L_{ij}$ tensor and the Weyl tensor in the
decomposition of the curvature:
\begin{eqnarray}
  \nabla_i T_j{}^k &=& \partial_i T_j{}^k - \Gamma_{ij}{}^\ell  \, T_\ell {}^k
  + \Gamma_{i\ell }{}^k \, T_j{}^\ell  \,, \\ {\mathbf +}R_{ijk}{}^\ell  &=&
  \partial_i \Gamma_{jk}{}^\ell  - \partial_j \Gamma_{ik}{}^\ell  +
  \Gamma_{im}{}^\ell  \, \Gamma_{jk}{}^m - \Gamma_{jm}{}^\ell  \,
  \Gamma_{ik}{}^m \,,\\ {\mathbf +}{\rm Ric}_{jk} &=& R_{ijk}{}^i
  \,,\\ {\mathbf +}R &=& R_{ij}{}^{ji}\,,\\ {\rm Weyl}_{ijk\ell } &=&
  R_{ijk\ell }\, {\mathbf +}\, \frac{4}{n-2}\,g_{[i|[k}L_{\ell]|j ]}\,.
\end{eqnarray}
An extensive comparison between the various conventions can be found
in \cite{mtw}.

\subsection{Variation of the Chern-Simons Lagrangian}
\label{vari}

We consider the Lagrangian
\begin{equation}
C_{\rm RR} = -\frac{1}{2} \, \left( \Gamma_\alpha{}^\beta \wedge
d\Gamma_\beta{}^\alpha - \frac{2}{3} \, \Gamma_\alpha{}^\beta \wedge
\Gamma_\beta{}^\gamma \wedge \Gamma_\gamma{}^\alpha \right)\,.
\end{equation}
The variation of this Chern-Simons Lagrangian, which only depends on the
connection, turns out to be
\begin{equation}
\delta C_{\rm RR} = -\delta\Gamma_\alpha{}^\beta \wedge R_\beta{}^\alpha
+\frac{1}{2}\,d\left(\Gamma_\alpha{}^\beta\wedge\delta\Gamma_\beta{}^\alpha
\right)\,.
\end{equation}
In the next step, we enforce vanishing torsion and nonmetricity by means of
respective Lagrange multiplier terms:
\begin{equation}
L=C_{\rm RR}
+ \lambda_\alpha \wedge T^\alpha + \lambda^{\alpha\beta}
\wedge Q_{\alpha\beta}\,.
\end{equation}
The variation then yields
\begin{eqnarray}
\delta L &=&
\delta C_{\rm RR} + \delta \lambda_\alpha \wedge T^\alpha
+ \lambda_\alpha \wedge \delta T^\alpha
+ \delta\lambda^{\alpha\beta} \wedge Q_{\alpha\beta}
+ \lambda^{\alpha\beta} \wedge \delta Q_{\alpha\beta}
\nonumber \\
&=&
-\delta \Gamma_\alpha{}^\beta \wedge R_\beta{}^\alpha
+ \delta\lambda_\alpha \wedge T^\alpha
+ \lambda_\alpha \wedge \left(d\delta\vartheta^\alpha
+ \delta \Gamma_\beta{}^\alpha \wedge \vartheta^\beta
+ \Gamma_\beta{}^\alpha \wedge \delta \vartheta^\beta \right)
\nonumber \\
&&
+\delta\lambda^{\alpha\beta} \wedge
Q_{\alpha\beta} + \lambda^{\alpha\beta} \wedge \left(-d\delta
g_{\alpha\beta} + \delta \Gamma_\alpha{}^\gamma\, g_{\gamma\beta} +
\Gamma_\alpha{}^\gamma \, \delta g_{\gamma\beta} +\delta\Gamma_\beta{}^\gamma
\, g_{\alpha\gamma} + \Gamma_\beta{}^\gamma \, \delta g_{\alpha\gamma} \right)
\nonumber\\&&
+\frac{1}{2} \, d
\left(\Gamma_\alpha{}^\beta\wedge\delta\Gamma_\beta{}^\alpha\right)
\nonumber \\
&=&
-\delta\Gamma_\alpha{}^\beta \wedge R_\beta{}^\alpha +\delta\lambda_\alpha
\wedge T^\alpha + \delta\lambda^{\alpha\beta} \wedge Q_{\alpha\beta}
+\frac{1}{2} \, d\left( \Gamma_\alpha{}^\beta \wedge
\delta\Gamma_\beta{}^\alpha \right) + \lambda_\alpha \wedge
D\delta\vartheta^\alpha
\nonumber \\
&&
- \delta\Gamma_\beta{}^\alpha \wedge \lambda_\alpha
\wedge \vartheta^\beta
-\lambda^{\alpha\beta} \wedge D\delta g_{\alpha\beta} +
\delta\Gamma_\alpha{}^\beta \wedge \left(\lambda^\alpha{}_\beta +
\lambda_\beta{}^\alpha \right)
\nonumber \\
&=&
\delta\lambda_\alpha \wedge T^\alpha + \delta\lambda^{\alpha\beta} \wedge
Q_{\alpha\beta}
+ \delta\vartheta^\alpha \wedge D\lambda_\alpha
+\delta g_{\alpha\beta} \, D\lambda^{\alpha\beta}
\nonumber \\
&&
-\delta\Gamma_\alpha{}^\beta \wedge \left(R_\beta{}^\alpha +
\lambda_\beta\wedge \vartheta^\alpha - \lambda^\alpha{}_\beta -
\lambda_\beta{}^\alpha \right)
\nonumber\\
&&
- d\left(-\lambda_\alpha \wedge \delta\vartheta^\alpha
+\frac{1}{2}
\Gamma_\alpha{}^\beta \wedge \delta \Gamma_\beta{}^\alpha -
\lambda^{\alpha\beta} \, \delta g_{\alpha\beta} \right) \,.
\end{eqnarray}

\subsection{Proof of eq.(\ref{dd*c})}\label{appendixbach}
By means of the Ricci identity and the decomposition of the curvature
we have
\begin{eqnarray}
  DD{}^\star C_\alpha &=& -R_\alpha{}^\beta \wedge {}^\star C_\beta
  \nonumber \\ &=& -{\rm Weyl}_\alpha{}^\beta \wedge {}^\star C_\beta
  -{{\rm Ricci}\!\!\!\!\!\!\!\!\nearrow}\,\,_{\alpha}{}^{\beta} \wedge
  {}^\star C_\beta -{\rm Scalar}_\alpha{}^\beta \wedge {}^\star
  C_\beta \,.
\end{eqnarray}
For $p$-forms $\phi$, $\psi$ of the same degree, there holds
${}^\star\phi\wedge\psi = {}^\star\psi\wedge\phi$. By means of
$\vartheta^\alpha \wedge {}^\star \phi = (-1)^{p-1} \,
{}^\star(e^\alpha \rfloor \phi)$ we can prove that ${\rm
  Scalar}_{\alpha\beta}\wedge{}^\star C^\alpha=0$.  Performing a
``partial integration'' we arrive at
\begin{equation}\label{texas}
DD{}^\star C_\alpha = -D\left({}^\star {\rm Weyl}_\alpha{}^\beta \wedge
L_\beta\right) + \left(D{}^\star {\rm Weyl}_\alpha{}^\beta\right) \wedge
L_\beta -
{}^\star{{\rm Ricci}\!\!\!\!\!\!\!\!\nearrow}\,\,_{\alpha}{}^{\beta}
\wedge C_\beta \,.
\end{equation}
Next, we use the ``double duality relations'' for the irreducible pieces of
the curvature,
\begin{eqnarray}
{}^\star {\rm Weyl}_{\alpha\beta}
&=& {\rm Weyl}_{\mu\nu} \, \frac{1}{2} \, \eta^{\mu\nu}{}_{\alpha\beta}
\,,\\
{}^\star {{\rm Ricci}\!\!\!\!\!\!\!\!\nearrow}\,\,_{\alpha\beta}
&=& - {{\rm Ricci}\!\!\!\!\!\!\!\!\nearrow}\,\,_{\mu\nu}
\, \frac{1}{2} \, \eta^{\mu\nu}{}_{\alpha\beta} \,,\\
{}^\star {\rm Scalar}_{\alpha\beta}
&=&
{\rm Scalar}_{\mu\nu} \, \, \frac{1}{2} \,
\eta^{\mu\nu}{}_{\alpha\beta} \,.
\end{eqnarray}
Together with eqs.(\ref{2ndbia2}) and (\ref{ldef1}), we obtain
\begin{eqnarray}
\left(D{}^\star{\rm Weyl}_\alpha{}^\beta\right) \wedge L_\beta
&=&
\frac{1}{2} \, \eta^{\mu\nu}{}_{\alpha}{}^{\beta} \, D{\rm Weyl}_{\mu\nu}
\wedge L_\beta
=-\, \frac{1}{2} \, \eta^{\mu\nu}{}_{\alpha}{}^{\beta} \,
\vartheta_{[\mu}\wedge C_{\nu]} \wedge L_\beta \nonumber \\
&=&
-\frac{1}{2} \, \eta^{\mu\beta\nu}{}_\alpha \,
\vartheta_{\mu}\wedge L_\beta\wedge C_\nu
=
-{}^\star
{{\rm Ricci}\!\!\!\!\!\!\!\!\nearrow}\,\,{}^\nu{}_\alpha \wedge C_\nu
+{}^\star
{\rm Scalar}^\nu{}_\alpha\wedge C_\nu
\nonumber\\
&=& {}^\star
{{\rm Ricci}\!\!\!\!\!\!\!\!\nearrow}\,\,{}_\alpha{}^\nu\wedge C_\nu \,.
\end{eqnarray}
Substituting this into (\ref{texas}) completes the proof.

\subsection{Computer algebra}\label{computeralgebra}
We present a Reduce-Excalc program for the computation of the Cotton
2-form in 3 dimensions and for testing of the DJT-field equations. The
components $C_\alpha{}^\beta$ are also assigned to a $3\times 3$
matrix. This matrix is then treated by the appropriate linear algebra
packages, as, for instance, the Reduce package ``Normform'' which
allows the determination of the Jordan form.  For an introduction into
the use of Reduce-Excalc in gravity theories see \cite{smh}.
\begin{verbatim}
% file cotton.exi chh 2003-08-08
% Calculation of Cotton tensor form given coframe/metric, n=3

load excalc ;

% Definition of coframe/metric
% eq.(4.7) in Percacci et al., Ann Phys (NY) 176 (1987) 344

coframe o(0) =  exp(mu*x/3)*(d t + 2*exp(-mu*x/3) * d y),
        o(1) =  d x,
        o(2) =  d y
with metric g = o(0)*o(0)-o(1)*o(1)-o(2)*o(2) ;

frame e ;

% calculation of curvature

pform riem2(a,b) = 2;
riemannconx chris1 ;
chris1(a,b) := chris1(b,a) ;
riem2(-a,b) := d chris1(-a,b) - chris1(-a,c) ^ chris1(-c,b) ;

% calculation of L_a and Cotton

pform ll1(a)=1,cotton2(a)=2;
ll1(a) :=    e(-b) _| riem2(a,b)
- 1/4 * (e(-c) _| ( e(-d) _| riem2(c,d) )) * o(a) ;

cotton2(a) := d ll1(a) + chris1(-b,a) ^ ll1(b) ;

% Definition of Cotton tensor
pform cotmat(a,b) = 0 ;
cotmat(a,b) :=  #(cotton2(a)^o(b)) ;

% Definitio of Cotton matrix
matrix cotm(3,3) $
for a:= 1:3 do {for b:= 1:3 do{
        cotm(a,b) := cotmat(-(a-1),b-1) }}$

% Definition Einstein 2-form
pform einstein2(a) =2;
einstein2(a) := (1/2) * #(o(a)^o(b)^o(c)) ^ riem2(-b,-c);

% Test of DJT field equations
pform null(a)=2;
null(a) := einstein2(a)+(1/mu)*cotton2(a);
\end{verbatim}

\end{document}